# Firehose instability in the space plasma with anisotropic Cairns-distribution electrons

Huo Rui, Du Jiulin

*Department of Physics, School of Science, Tianjin University, Tianjin 300350, China*

**Abstract** We study the electron firehose mode propagating parallel to the ambient magnetic field in the space plasma with anisotropic Cairns-distribution electrons. The dispersion relation, the wave frequency $\omega_r/\Omega_i$ and the growth rate $\gamma/\Omega_i$ of electron firehose mode are derived, and the condition for onset of the firehose instability is obtained. We show that the wave frequency and the growth rate both depend significantly on the parameters, such as the parallel electron beta $\beta_{e\|}$, the nonthermal parameter $\Lambda$ and the electron temperature anisotropy $A_e$, and the anisotropic Cairns-distribution electrons change the instability condition. The numerical analyses show that the wave frequency and the growth rate of the electron firehose mode increase with increase of the parameters. The results may be helpful for understanding the firehose instability in space plasma environments.

**Keywords:** Electron firehose mode, Cairns-distribution, Temperature anisotropy, Wave and instability

## 1. Introduction

Space plasmas are usually observed to be low-collisional, or even to be collisionless, therefore, the interaction between waves and particles, as well as the behavior of particles under electromagnetic fluctuations, have become the main dynamic mechanisms [1, 2]. The existence of these fluctuations has been confirmed by experimental observation and found to be closely related to kinetic instabilities [3-5]. Temperature anisotropies of plasma particle velocity distributions are the origin of various kinetic instabilities that can enhance the fluctuations [6, 7]. The electron firehose instability is caused by temperature anisotropy $T_{e\|}>T_{e\perp}$ for electron populations [8-10], where $\|$ and $\perp$ denote the directions parallel and perpendicular to the background magnetic field, respectively. The electron firehose instability reduces the increase of the electron-temperature anisotropy and could be one of the most efficient mechanisms of temperature isotropy in solar flares [11]. Hollweg and Volk investigated the electron firehose mode propagating parallel to the background magnetic field in a hot plasma caused by the temperature anisotropy associated with electrons [12]. The electron firehose modes can resonate with ions only when the electron anisotropy is not very large and therefore the energy of electrons can be transferred to protons in the solar wind. However, these modes will also resonate with electrons if the electron temperature in the parallel direction is much greater than that in the perpendicular direction [13]. Gary and Madland studied the maximum growth rate of the electron firehose modes for parallel propagation [14]. Paesold and Benz investigated the electron firehose modes for oblique propagation and found that the maximum growth rate of these modes for oblique propagation is greater than that for parallel propagation [15]. Recently, the electron firehose mode instability has been discussed by taking different limits of plasma and temperature anisotropy [16-19].

In space plasmas, most particles always show deviations from thermodynamic equilibrium and have high energy tails [20-23]. There are many non-Maxwellian velocity distribution functions to describe these high-energy particles, such as nonextensive q-distribution [24], the kappa distribution [25], the (*r*, *q*) distribution [26], the nonthermal Cairns-distribution [27], Mixed Cairns-Tsallis distribution [28], so on. Cairns et al. introduced a distribution function to study the ion solitary structures observed in the upper ionosphere with non-thermal electrons, which is given

as [29]

$$f_e\left(v_\perp, v_\parallel\right) = \frac{1}{\pi^{3/2}\left(1+\frac{11}{4}\Lambda\right)v_{te\perp}^2 v_{te\parallel}}\left[1+\Lambda\left(\frac{v_\perp^4}{v_{te\perp}^4}+\frac{v_\parallel^4}{v_{te\parallel}^4}\right)\right]\exp\left(-\frac{v_\perp^2}{v_{te\perp}^2}-\frac{v_\parallel^2}{v_{te\parallel}^2}\right) \quad (1)$$

where $v_{te\perp} = \sqrt{2T_{e\perp}/m_e}$ and $v_{te} = \sqrt{2T_e/m_e}$ are the perpendicular and parallel thermal velocities of the electrons relative to the magnetic field direction, respectively. $\Lambda$ is the non-thermality parameter, which indicates the population of non-thermal electrons. The Cairns-distribution has been widely used in the theoretical study of plasma waves [30]. Hadi *et al* investigated the effect of the Cairns-distribution on Landau damping of electrostatic modes [31]. Usman *et al* studied parallel whistler instability in a plasma with anisotropic Cairns-distribution [32]. Azra *et al* discussed the ordinary mode instability in a plasma with non-thermal electrons [33]. However, a survey of the literature shows that no one has yet used the Cairns-distribution to investigate electron firehose instability. In this paper, we study the electron firehose instability with anisotropic Cairns-distribution.

The paper is organized as follows: In section 2, we present the plasma model to derive analytically the dispersion relation of the electron firehose mode with temperature anisotropy of Cairns-distributed electrons, and then we get expressions for the wave frequency and growth rate. In section 3, we give numerical analyses of the frequency and growth rate of electron firehose mode. In section 4, we present a summary.

## 2. The firehose mode with anisotropic Cairns-distribution electrons

We now introduce the theory of electromagnetic waves for the homogeneous and collisionless plasma. The Vlasov-Maxwell equations that the system satisfies can be expressed as [34]:

$$\left[\frac{\partial}{\partial t}+\mathbf{v}\cdot\frac{\partial}{\partial \mathbf{r}}+\frac{q_\alpha}{m_\alpha}\left(\mathbf{E}+\frac{\mathbf{v}\times\mathbf{B}}{c}\right)\cdot\frac{\partial}{\partial \mathbf{v}}\right]f_\alpha(\mathbf{r},\mathbf{v},t) = 0, \quad (2)$$

$$\nabla\times\mathbf{E} = -\frac{1}{c}\frac{\partial \mathbf{B}}{\partial t}, \quad (3)$$

$$\nabla\times\mathbf{B} = \frac{4\pi\mathbf{J}}{c}+\frac{1}{c}\frac{\partial \mathbf{E}}{\partial t}, \quad (4)$$

where $f_\alpha(\mathbf{r}, \mathbf{v}, t)$ is the velocity distribution function of particles. $\mathbf{E}$ and $\mathbf{B}$ are the electric field intensity and the magnetic induction, respectively; $q_\alpha$ and $m_\alpha$ are respectively the charge and mass of the particle, and $c$ is the light speed. We adopt the gyro-coordinates in the $\mathbf{v}$-space with its z-axis parallel to $\mathbf{B} = B_0\hat{\mathbf{z}}$ assume that the direction of wave vector $\mathbf{k}$ to be along the z-axis. By linearizing and solving the above equations, the dispersion relation for the left-hand and right-hand circularly polarized waves propagating parallel to the ambient magnetic field can expressed as

$$D(\omega,k) = 1-\frac{c^2k^2}{\omega^2}+\frac{1}{2}\sum_\alpha\frac{\omega_{p\alpha}^2}{\omega^2}\int d\mathbf{v}\frac{v_\perp F(\omega,v_\perp,v_\parallel)}{\omega-k_\parallel v_\parallel-\Omega_\alpha} = 0, \quad (5)$$

$$D(\omega,k) = 1-\frac{c^2k^2}{\omega^2}+\frac{1}{2}\sum_\alpha\frac{\omega_{p\alpha}^2}{\omega^2}\int d\mathbf{v}\frac{v_\perp F(\omega,v_\perp,v_\parallel)}{\omega-k_\parallel v_\parallel+\Omega_\alpha} = 0, \quad (6)$$

where $\Omega_\alpha = q_\alpha B_0/m_\alpha c$ is the gyro-frequency of the particle and we have used the abbreviation,

$$F(\omega, v_\perp, v_\parallel) \equiv (\omega - k_\parallel v_\parallel) \frac{\partial f_{\alpha 0}}{\partial v_\perp} + k_\parallel v_\perp \frac{\partial f_{\alpha 0}}{\partial v_\parallel}. \tag{7}$$

In order to derive the dispersion relation for the electron firehose mode, for simplicity, we consider an electron-ion plasma in which the electrons are described by anisotropic Cairns-distribution in Eq. (1) and the ions are described by isotropic Maxwellian distribution as follows:

$$f_i(v) = \frac{1}{\pi^{3/2} v_{ti}^3} \exp\left(-\frac{v^2}{v_{ti}^2}\right), \tag{8}$$

where $v_{ti} = \sqrt{2T_i/m_i}$ are the thermal velocities of the ions. The electron firehose instability is left-hand circularly polarized. Taking Eqs. (1) and (8) into Eq.(5), we can derive (see Appendix A) that

$$D(\omega, k) = 1 - \frac{k^2 c^2}{\omega^2} - \frac{1}{\pi^{1/2}} \frac{\omega_{pi}^2}{k v_{ti} \omega} \int_{-\infty}^{\infty} d\eta_i \frac{\exp(-\eta_i^2)}{\xi_i - \eta_i} - \frac{4}{\pi^{1/2}(4+11\Lambda)} \frac{\omega_{pe}^2}{k\omega v_{te\parallel}} \times$$

$$\int_{-\infty}^{\infty} d\eta_e \frac{\exp(-\eta_e^2)}{\xi_e - \eta_e} \left\{ 1 + 2\Lambda + \Lambda \eta_e^4 - \frac{k v_{te\parallel}}{\omega} \eta_e \left[ 1 + 2\Lambda + \Lambda \eta_e^4 - \frac{v_{te\perp}^2}{v_{te\parallel}^2} \left(1 + 6\Lambda + 2\Lambda \eta_e^2 - \Lambda \eta_e^4\right) \right] \right\} = 0, \tag{9}$$

where $\xi_\alpha = \omega - \Omega_\alpha / k v_{t\alpha}$, $\eta_\alpha = v/v_{t\alpha}$. For the electron firehose mode, the electrons are nonresonant ($\xi_e \gg 1$) and the ions are resonant ($\xi_i < 1$). In this case, we can calculate the Eq. (9) to obtain that (see Appendix B)

$$D(\omega, k) = 1 - \frac{k^2 c^2}{\omega^2} - \frac{\omega_{pe}^2}{\omega} \frac{1}{\omega - \Omega_e}$$
$$+ \frac{2}{4+11\Lambda} \frac{\omega_{pe}^2}{\omega^2} \frac{(k v_{te\parallel})^2}{(\omega - \Omega_e)^2} \left[ A_e + \Lambda \left( \frac{27}{4} A_e - 1 \right) \right] + i\sqrt{\pi} \frac{\omega_{pi}^2}{k v_{ti} \omega} \exp(-\xi_i^2) = 0, \tag{10}$$

where $A_e = 1 - T_{e\perp}/T_e$. Inserting $\omega = \omega_r + i\gamma$ into Eq. (10), in the case of $\gamma \ll \omega_r$, then we can get

$$\text{Re} D(\omega_r, k) = 1 - \frac{k^2 c^2}{\omega_r^2} - \frac{\omega_{pe}^2}{\omega_r} \frac{1}{\omega_r - \Omega_e} + \frac{2}{4+11\Lambda} \frac{\omega_{pe}^2}{\omega_r^2} \frac{(k v_{te\parallel})^2}{(\omega_r - \Omega_e)^2} \left[ A_e + \Lambda \left( \frac{27}{4} A_e - 1 \right) \right], \tag{11}$$

$$\text{Im} D(\omega_r, k) = \sqrt{\pi} \frac{\omega_{pi}^2}{k v_{ti} \omega_r} \exp(-\xi_i^2). \tag{12}$$

Making $\text{Re} D(\omega_r, k) = 0$, we can get an equation for the wave frequency of the electron firehose mode,

$$1 - \frac{k^2 c^2}{\omega_r^2} - \frac{\omega_{pe}^2}{\omega_r} \frac{1}{\omega_r - \Omega_e} + \frac{2}{4+11\Lambda} \frac{\omega_{pe}^2}{\omega_r^2} \frac{(k v_{te\parallel})^2}{(\omega_r - \Omega_e)^2} \left[ A_e + \Lambda \left( \frac{27}{4} A_e - 1 \right) \right] = 0. \tag{13}$$

Because $\omega^2 \ll c^2 k^2$, $\gamma < \omega_r \ll |\Omega_e|$, the unity in Eq. (13) can be ignored. Multiplying both sides by $\omega_r^2 / \omega_{pi}^2$, then we get that

$$-\frac{k^2 c^2}{\omega_{pi}^2} - \frac{\omega_{pe}^2}{\omega_{pi}^2} \frac{\omega_r}{|\Omega_e|} + \frac{2}{4+11\Lambda} \frac{\omega_{pe}^2}{\omega_{pi}^2} \frac{(k v_{te\parallel})^2}{\Omega_e^2} \left[ A_e + \Lambda \left( \frac{27}{4} A_e - 1 \right) \right] = 0. \tag{14}$$

By using the formula $\frac{\omega_{pe}^2}{\omega_{pi}^2} \Omega_i = |\Omega_e|$ and $\frac{\omega_{pe}^2}{c^2} \frac{v_{te\parallel}^2}{\Omega_e^2} = \frac{4\pi n_0 e^2}{c^2 m_e} \frac{2T_e}{m_e} \frac{m_e^2 c^2}{e^2 B^2} = \frac{8\pi n_0 T_e}{B^2} = \beta_{e\parallel}$, we can obtain the wave frequency for the electron firehose mode,

$$\frac{\omega_r}{\Omega_i} = -\frac{k^2 c^2}{\omega_{pi}^2}\left\{1 - \frac{2\beta_{e\|}}{4+11\Lambda}\left[A_e + \Lambda\left(\frac{27}{4}A_e - 1\right)\right]\right\}. \tag{15}$$

Using the expression $\gamma = -\dfrac{\operatorname{Im} D(\omega_r, k)}{\partial \operatorname{Re} D(\omega_r, k)/\partial \omega_r}$, we can get that

$$\frac{\partial \operatorname{Re} D(\omega_r, k)}{\partial \omega_r} = \frac{2}{\omega_r^3} k^2 c^2 \left\{1 - \frac{2\beta_{e\|}}{4+11\Lambda}\left[A_e + \Lambda\left(\frac{27}{4}A_e - 1\right)\right]\right\} + \frac{\omega_{pe}^2}{\omega_r^2 |\Omega_e|} \tag{16}$$

and the growth rate for electron firehose mode with anisotropic Cairns-distribution electrons,

$$\frac{\gamma}{\Omega_i} = -\sqrt{\pi}\frac{\Omega_i}{kv_{ti}}\frac{k^2 c^2}{\omega_{pi}^2}\left\{1 - \frac{2\beta_{e\|}}{4+11\Lambda}\left[A_e + \Lambda\left(\frac{27}{4}A_e - 1\right)\right]\right\}\exp(-\xi_i^2). \tag{17}$$

The electron firehose instability in the Cairns-distributed plasma will occur if the condition

$$1 - \frac{2\beta_{e\|}}{4+11\Lambda}\left[A_e + \Lambda\left(\frac{27}{4}A_e - 1\right)\right] < 0 \tag{18}$$

is satisfied. In the limit $\Lambda = 0$, it becomes the condition of the firehose instability for the plasma with a bi-Maxwellian distribution [12],

$$1 - \frac{\beta_{e\|} A_e}{2} < 0. \tag{19}$$

### 3. Numerical analyses

In this section, we make numerical analyses of the dispersion relation in Eq. (15) and the damping rate in Eq. (17) for the electron firehose mode in the space plasma with anisotropic Cairns-distribution electrons. The main purpose of this study is to show the effect of nonthermal parameter $\Lambda$ on the electron firehose mode. Besides, the effects of some other plasma parameters on the electron firehose mode have also been discussed. In the numerical analyses, we used the parameters appropriate to the solar wind plasma [35, 36]: $\beta_{e\|} \geq 3$, $A_e \sim 0.5$, and $\Omega_i/kv_{ti} \sim 0.01$.

Figure 1 describes the variation of the wave frequency $\omega_r/\Omega_i$ as a function of $kc/\omega_{pi}$ for the electron firehose mode with different parameters. Fig. 1(a) shows the effect of nonthermal parameter $\Lambda$ on the wave frequency with fixed plasma parameters $\beta_{e\|} = 5$, $A_e = 0.5$. The line with $\Lambda = 0$ corresponds to the Maxwellian case of the plasma, the variation of wave frequency $\omega_r/\Omega_i$ with $kc/\omega_{pi}$ is consistent with the result of bi-Maxwellian distribution [15]. It is found that the wave frequency will increase as the nonthermal parameter $\Lambda$ increases. Therefore, the wave frequency for electron firehose mode in the plasma with the Cairns-distribution electrons is generally greater than with the Maxwellian distribution electrons. Fig. 1(b) depicts the effect of electron plasma beta $\beta_{e\|}$ on the wave frequency with fixed plasma parameters $\Lambda = 0.2$, $A_e = 0.5$. It is observed that the larger parameter $\beta_{e\|}$ will lead to the larger wave frequency for electron firehose modes. Fig. 1(c) exhibits the variation of the wave frequency with electron temperature anisotropy parameter $A_e$ with fixed plasma parameters $\Lambda = 0.2$, $\beta_{e\|} = 5$. It is concluded that the wave frequency for electron firehose modes will increase with the increase of $A_e$.

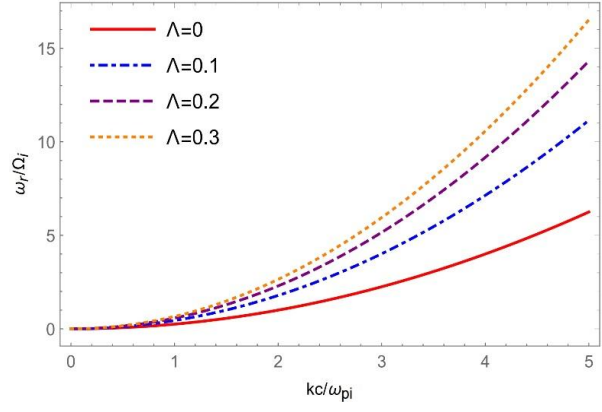

(a) $\beta_{e\|} = 5$, $A_e = 0.5$

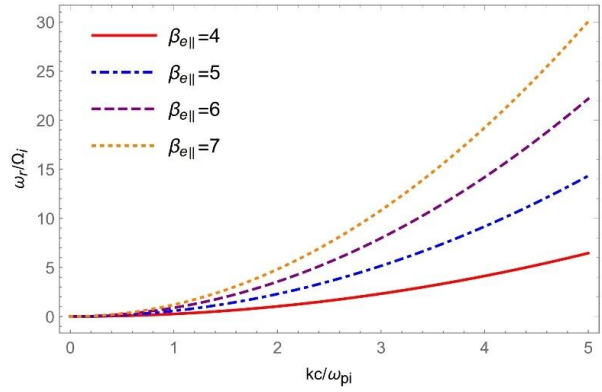

(b) $\Lambda = 0.2$, $A_e = 0.5$

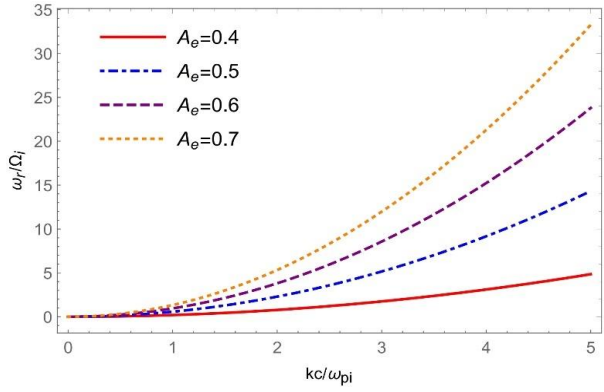

(c) $\Lambda = 0.2$, $\beta_{e\|} = 5$

**Figure 1.** $\omega_r / \Omega_i$ as a function of $kc / \omega_{pi}$ for the electron firehose mode

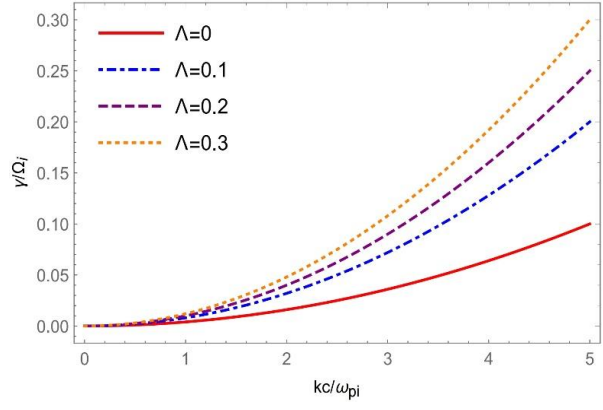

(a) $\beta_{e\parallel} = 5$, $A_e = 0.5$

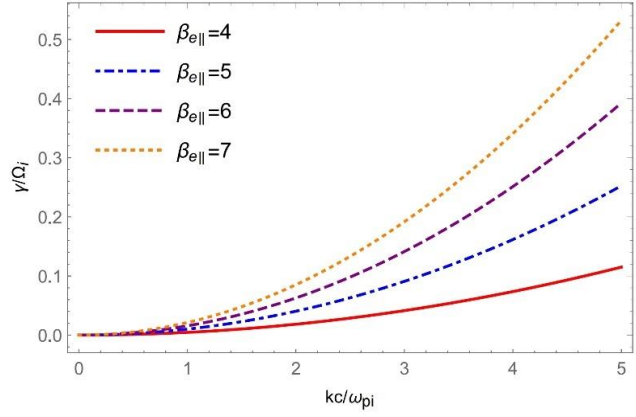

(b) $\Lambda = 0.2$, $A_e = 0.5$

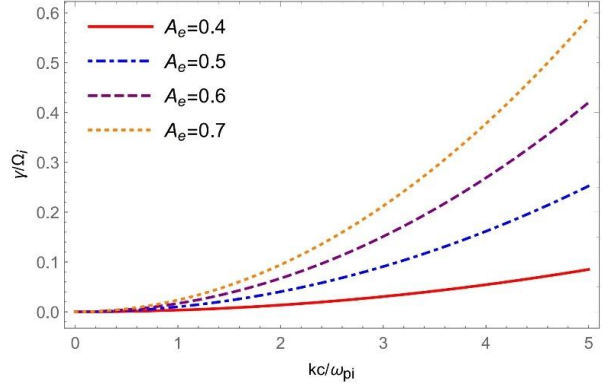

(c) $\Lambda = 0.2$, $\beta_{e\parallel} = 5$

**Figure 2.** $\gamma/\Omega_i$ as a function of $kc/\omega_{pi}$ for the electron firehose mode

Figure 2 describes the variation of the growth rate $\gamma/\Omega_i$ as a function of $kc/\omega_{pi}$ for electron firehose mode with different parameters. Fig. 2(a) shows the effect of nonthermal parameter $\Lambda$ on the growth rate with fixed plasma parameters $\beta_{e\parallel} = 5$, $A_e = 0.5$. The line with $\Lambda = 0$ corresponds to the Maxwellian case of the plasma. It is found that the growth rate of the wave is enhanced by increasing the nonthermal parameter $\Lambda$. This rise in growth rate at bigger values of $\Lambda$ reveals the availability of more free energy to excite the electron firehose instability. Fig. 2(b) depicts the effect of parallel electron beta $\beta_{e\parallel}$ on the growth rate with fixed plasma parameters $\Lambda = 0.2$, $A_e = 0.5$. It can be concluded that the growth rate increases for higher parallel electron beta. This also means that higher values of the parallel electron beta can make more contributions to the availability of more free energy to make the mode unstable in a direction parallel to the ambient magnetic field. Fig. 2(c) gives the effect of electron temperature anisotropy parameter $A_e$ on the growth rate with fixed plasma parameters $\Lambda = 0.2$, $\beta_{e\parallel} = 5$. It is obvious that the growth rate increases with the increase of electron temperature anisotropy $A_e$. Therefore, higher values of electron temperature anisotropy also play a key role in increasing free energy for the electron firehose mode.

## 4. Conclusions

In this paper, by employing the kinetic theory, the wave frequency $\omega_r / \Omega_i$ in Eq.(15) and the damping rate $\gamma / \Omega_i$ in Eq. (17) for electron firehose mode with anisotropic Cairns distribution electrons are investigated. The condition (18) for the onset of the electron firehose instability with Cairns distribution is also studied. It is found that the wave frequency in Eq. (15) and the damping rate in Eq. (17) for electron firehose mode depend on the parallel electron beta $\beta_{e\parallel}$ (including plasma density, magnetic field), nonthermal parameter $\Lambda$, electron temperature anisotropy $A_e$. The wave frequency for electron firehose mode is larger in the plasma with anisotropic Cairns-distribution electrons than that in the Maxwellian case. Besides, the wave frequency is also enhanced by increasing the parallel electron beta $\beta_{e\parallel}$ and electron temperature anisotropy $A_e$. It is observed that the growth rate for electron firehose mode is larger with the higher values of parallel electron beta $\beta_{e\parallel}$, nonthermal parameter $\Lambda$, electron temperature anisotropy $A_e$. The physical reason for this variation in growth is that the higher values of those can make contributions to the availability of more free energy to excite the electron firehose instability. These results may be helpful for the study of wave emission, particle acceleration, and turbulence in space plasmas.

**Appendix A**

For Eq. (9), taking the velocity distribution of electrons in Eq. (1) and the velocity distribution of ions in Eq. (8) into Eq. (5), using $d^3v = 2\pi v_\perp dv_\perp dv_\parallel$, we can get that

$$1 - \frac{k^2c^2}{\omega^2} + \frac{\pi}{\omega^2}\sum_\alpha \omega_{p\alpha}^2 \int_{-\infty}^{\infty} \frac{dv_\parallel}{\omega - \Omega_\alpha - kv_\parallel} \int_0^\infty v_\perp^2 F(\omega, v_\perp, v_\parallel) dv_\perp = 0, \quad (A1)$$

where

$$F(\omega, v_\perp, v_\parallel) \equiv \left(\omega - k_\parallel v_\parallel\right)\frac{\partial f_{\alpha 0}}{\partial v_\perp} + k_\parallel v_\perp \frac{\partial f_{\alpha 0}}{\partial v_\parallel}. \quad (A2)$$

For electrons, we have that

$$\frac{\partial f_0}{\partial v_\perp} = \frac{2v_\perp}{\pi^{3/2}\left(1+\frac{11}{4}\Lambda\right)v_{te\perp}^4 v_{te\parallel}} \left\{2\Lambda\frac{v_\perp^2}{v_{te\perp}^2} - \left[1 + \Lambda\left(\frac{v_\perp^4}{v_{te\perp}^4} + \frac{v_\parallel^4}{v_{te\parallel}^4}\right)\right]\right\} \exp\left(-\frac{v_\perp^2}{v_{te\perp}^2} - \frac{v_\parallel^2}{v_{te\parallel}^2}\right), \quad (A3)$$

$$\frac{\partial f_0}{\partial v_\parallel} = \frac{2v_\parallel}{\pi^{3/2}\left(1+\frac{11}{4}\Lambda\right)v_{te\perp}^2 v_{te\parallel}^3} \left[2\Lambda\frac{v_\parallel^2}{v_{te\parallel}^2} - 1 - \Lambda\left(\frac{v_\perp^4}{v_{te\perp}^4} + \frac{v_\parallel^4}{v_{te\parallel}^4}\right)\right] \exp\left(-\frac{v_\perp^2}{v_{te\perp}^2} - \frac{v_\parallel^2}{v_{te\parallel}^2}\right), \quad (A4)$$

$$\frac{\pi}{\omega}\sum_{\alpha}\omega_{p\alpha}^{2}\int_{-\infty}^{\infty}\frac{dv_{\parallel}}{\omega-\Omega_{\alpha}-kv_{\parallel}}\int_{0}^{\infty}v_{\perp}^{2}\left[\left(1-\frac{kv_{\parallel}}{\omega}\right)\frac{\partial f_{\alpha}}{\partial v_{\perp}}+\frac{kv_{\perp}}{\omega}\frac{\partial f_{\alpha}}{\partial v_{\parallel}}\right]dv_{\perp}$$

$$=-\frac{\omega_{pe}^{2}}{\pi^{3/2}\left(1+\frac{11}{4}\Lambda\right)v_{te\parallel}}\frac{\pi}{\omega}\int_{-\infty}^{\infty}\frac{dv_{\parallel}}{\omega-\Omega_{e}-kv_{\parallel}}\left\{\begin{array}{l}\left(1-\frac{kv_{\parallel}}{\omega}\right)\left(\Lambda\frac{v_{\parallel}^{4}}{v_{te\parallel}^{4}}+2\Lambda+1\right)\\+\frac{kv_{\parallel}}{\omega}\frac{v_{te\perp}^{2}}{v_{te\parallel}^{2}}\left[1+\left(6-2\frac{v_{\parallel}^{2}}{v_{te\parallel}^{2}}+\frac{v_{\parallel}^{4}}{v_{te\parallel}^{4}}\right)\Lambda\right]\end{array}\right\}\exp\left(-\frac{v_{\parallel}^{2}}{v_{te\parallel}^{2}}\right)$$

$$=\frac{1}{\pi^{1/2}\left(1+\frac{11}{4}\Lambda\right)}\frac{\omega_{pe}^{2}}{k\omega v_{te\parallel}}\int_{-\infty}^{\infty}\frac{d\eta_{e}}{\xi-\eta_{e}}\left\{\begin{array}{l}\left(\frac{kv_{te\parallel}}{\omega}\eta_{e}-1\right)\left(\Lambda\eta_{e}^{4}+1+2\Lambda\right)\\-\frac{k}{\omega}\frac{v_{te\perp}^{2}}{v_{te\parallel}^{2}}\eta_{e}\left[1+\left(6-2\eta_{e}^{2}+\eta_{e}^{4}\right)\Lambda\right]\end{array}\right\}\exp\left(-\eta_{e}^{2}\right)$$

$$=-\frac{1}{\pi^{1/2}\left(1+\frac{11}{4}\Lambda\right)}\frac{\omega_{pe}^{2}}{k\omega v_{te\parallel}}\int_{-\infty}^{\infty}d\eta_{e}\frac{\exp\left(-\eta_{e}^{2}\right)}{\xi-\eta_{e}}\left\{1+2\Lambda-\eta_{e}\frac{kv_{te\parallel}}{\omega}\left[\begin{array}{l}1+2\Lambda-\Lambda\eta_{e}^{3}+\Lambda\eta_{e}^{4}\\+\frac{v_{te\perp}^{2}}{v_{te\parallel}^{2}}\left(2\Lambda\eta_{e}^{2}-\Lambda\eta_{e}^{4}-6\Lambda-1\right)\end{array}\right]\right\}, \quad (A5)$$

where $\xi_{e}=\frac{\omega-\Omega_{e}}{kv_{te}}$, $\eta_{e}=\frac{v}{v_{te}}$.

For ions, we have that

$$f_{i}(v)=\frac{1}{\pi^{3/2}v_{ti}^{3}}\exp\left(-\frac{v^{2}}{v_{ti}^{2}}\right),$$

$$\frac{\pi}{\omega}\omega_{pi}^{2}\int_{-\infty}^{\infty}\frac{dv_{\parallel}}{\omega-\Omega_{i}-kv_{\parallel}}\int_{0}^{\infty}v_{\perp}^{2}\left[\left(1-\frac{kv_{\parallel}}{\omega}\right)\frac{\partial f_{i}}{\partial v_{\perp}}+\frac{kv_{\perp}}{\omega}\frac{\partial f_{i}}{\partial v_{\parallel}}\right]dv_{\perp}=\frac{\pi}{\omega}\omega_{pi}^{2}\int_{-\infty}^{\infty}\frac{dv_{\parallel}}{\omega-\Omega_{i}-kv_{\parallel}}\int_{0}^{\infty}v_{\perp}^{2}\frac{\partial f_{i}}{\partial v_{\perp}}dv_{\perp}$$

$$=-\frac{1}{\pi^{1/2}v_{ti}}\frac{\omega_{pi}^{2}}{\omega}\int_{-\infty}^{\infty}\frac{dv_{\parallel}}{\omega-\Omega_{i}-kv_{\parallel}}\exp\left(-\frac{v_{\parallel}^{2}}{v_{ti}^{2}}\right)=-\frac{1}{\pi^{1/2}}\frac{\omega_{pi}^{2}}{kv_{ti}\omega}\int_{-\infty}^{\infty}\frac{dv_{\parallel}}{(\omega-\Omega_{i})\frac{1}{k}-v_{\parallel}}\exp\left(-\frac{v_{\parallel}^{2}}{v_{ti}^{2}}\right)$$

$$=\frac{1}{\pi^{1/2}}\frac{\omega_{pi}^{2}}{kv_{ti}\omega}\int_{-\infty}^{\infty}\frac{\exp\left(-\eta_{i}^{2}\right)}{\eta_{i}-\xi_{i}}d\eta_{i}, \quad (A6)$$

where $\xi_{i}=\frac{\omega-\Omega_{i}}{kv_{ti}}$, $\eta_{i}=\frac{v_{\parallel}}{v_{ti}}$. So we obtain that

$$D(\omega,k)=1-\frac{k^{2}c^{2}}{\omega^{2}}-\frac{1}{\pi^{1/2}}\frac{\omega_{pi}^{2}}{kv_{ti}\omega}\int_{-\infty}^{\infty}d\eta_{i}\frac{\exp\left(-\eta_{i}^{2}\right)}{\xi_{i}-\eta_{i}}-\frac{4}{\pi^{1/2}\left(4+11\Lambda\right)}\frac{\omega_{pe}^{2}}{k\omega v_{te\parallel}}\times$$

$$\int_{-\infty}^{\infty}d\eta_{e}\frac{\exp\left(-\eta_{e}^{2}\right)}{\xi_{e}-\eta_{e}}\left\{1+2\Lambda+\Lambda\eta_{e}^{4}-\frac{kv_{te\parallel}}{\omega}\eta_{e}\left[1+2\Lambda+\Lambda\eta_{e}^{4}-\frac{v_{te\perp}^{2}}{v_{te\parallel}^{2}}\left(1+6\Lambda+2\Lambda\eta_{e}^{2}-\Lambda\eta_{e}^{4}\right)\right]\right\}=0.$$

(A7)

**Appendix B**

$$D(\omega,k)=1-\frac{k^{2}c^{2}}{\omega^{2}}-\frac{1}{\pi^{1/2}}\frac{\omega_{pi}^{2}}{kv_{ti}\omega}\int_{-\infty}^{\infty}d\eta_{i}\frac{\exp\left(-\eta_{i}^{2}\right)}{\xi_{i}-\eta_{i}}-\frac{4}{\pi^{1/2}\left(4+11\Lambda\right)}\frac{\omega_{pe}^{2}}{k\omega v_{te\parallel}}\times$$

$$\int_{-\infty}^{\infty}d\eta_{e}\frac{\exp\left(-\eta_{e}^{2}\right)}{\xi_{e}-\eta_{e}}\left\{1+2\Lambda+\Lambda\eta_{e}^{4}-\eta_{e}\frac{kv_{te\parallel}}{\omega}\left[1+2\Lambda+\Lambda\eta_{e}^{4}-\frac{v_{te\perp}^{2}}{v_{te\parallel}^{2}}\left(1+6\Lambda-2\Lambda\eta_{e}^{2}+\Lambda\eta_{e}^{4}\right)\right]\right\}=0.$$

(B1)

For electrons, $\xi \gg 1$, the last term in Eq.(B1) is calculated as

$$\frac{4\omega_{pe}^2/(k\omega v_{te\parallel})}{\pi^{1/2}(4+11\Lambda)}\int_{-\infty}^{\infty}d\eta_e\left(\frac{1}{\xi}+\frac{\eta_e}{\xi^2}+\frac{\eta_e^2}{\xi^3}\right)\left\{1+2\Lambda+\Lambda\eta_e^4-\eta_e\frac{kv_{te\parallel}}{\omega}\left[1+2\Lambda+\Lambda\eta_e^4-\frac{v_{te\perp}^2}{v_{te\parallel}^2}\left(1+6\Lambda-2\Lambda\eta_e^2+\Lambda\eta_e^4\right)\right]\right\}\exp(-\eta_e^2)$$

$$=-\frac{4\omega_{pe}^2}{(4+11\Lambda)k\omega v_{te\parallel}}\left\{\frac{1+2\Lambda}{\xi}+\frac{3}{4}\frac{\Lambda}{\xi}+\frac{3}{2\xi^2}\frac{kv_{te\parallel}}{\omega}\left[\frac{v_{te\perp}^2}{v_{te\parallel}^2}\Lambda+\frac{5}{4}\Lambda\left(1-\frac{v_{te\perp}^2}{v_{te\parallel}^2}\right)+\frac{1}{3}\left(1+2\Lambda-\frac{v_{te\perp}^2}{v_{te\parallel}^2}(1+6\Lambda)\right)\right]\right\}$$

$$=-\frac{\omega_{pe}^2}{\omega}\frac{1}{\omega-\Omega_e}+\frac{2}{4+11\Lambda}\frac{\omega_{pe}^2}{\omega^2}\frac{(kv_{te\parallel})^2}{(\omega-\Omega_e)^2}\left[1+\frac{23}{4}\Lambda-\frac{v_{te\perp}^2}{v_{te\parallel}^2}\left(1+\frac{27}{4}\Lambda\right)\right]$$

$$=-\frac{\omega_{pe}^2}{\omega}\frac{1}{\omega-\Omega_e}+\frac{2}{4+11\Lambda}\frac{\omega_{pe}^2}{\omega^2}\frac{(kv_{te\parallel})^2}{(\omega-\Omega_e)^2}\left[A_e+\Lambda\left(\frac{27}{4}A_e-1\right)\right], \quad (B2)$$

where we used $A_e=1-T_{e\perp}/T_e$.

For ions, $\xi_i<1$, we have that

$$-\frac{1}{\pi^{1/2}}\frac{\omega_{pi}^2}{kv_{ti}\omega}\int_{-\infty}^{\infty}\frac{\exp(-y^2)}{\xi_i-y}dy=i\sqrt{\pi}\frac{\omega_{pi}^2}{kv_{ti}\omega}\exp(-\xi_i^2), \quad (B3)$$

therefore,

$$D(\omega,k)=1-\frac{k^2c^2}{\omega^2}-\frac{\omega_{pe}^2}{\omega}\frac{1}{\omega-\Omega_e}$$
$$+\frac{2}{4+11\Lambda}\frac{\omega_{pe}^2}{\omega^2}\frac{(kv_{te\parallel})^2}{(\omega-\Omega_e)^2}\left[A_e+\Lambda\left(\frac{27}{4}A_e-1\right)\right]+i\sqrt{\pi}\frac{\omega_{pi}^2}{kv_{ti}\omega}\exp(-\xi_i^2)=0. \quad (B4)$$

**Data Availability**

Data sharing is not applicable to this article as no new data were created or analyzed in this study.